# Ultra-high-Q nanobeam cavity design in Diamond

Igal Bayn[1*], Joseph Salzman[1] and Rafi Kalish[2]

*1 Department of Electrical Engineering and Microelectronics Research Center, Technion Haifa, 32000, Israel*
*2 Department of Physics and Solid State Institute, Technion Haifa, 32000, Israel*

*Corresponding author: eebayn@techunix.technion.ac.il*

**Abstract:** A novel nanobeam design with a triangular cross-section is proposed. This design makes possible implementing nanocavities with improved optical properties. The dependence of a diamond-based cavity quality factor $Q$ and mode volume $V_m$ on geometry parameter space are studied via 3D FDTD computations. An ultra-high-$Q$ cavity with $Q \approx 2.51 \times 10^6$ and $V_m = 1.06 \times (\lambda/n)^3$ is predicted. The mode preferential radiation is upward. The implications on the potential applications are discussed. The proposed nanobeam enables fabrication of the cavity without relying on a pre-existing free-standing diamond membrane as required in most previous approaches.

## 1. Introduction

Quantum spintronics is a most promising field in future applications, such as quantum information processing, quantum computations and nano-bio sensing. It relies on the luminescence of particular single defect states (qubits) which can be initialized into a two state superposition (ground and excited). At the moment the most promising candidate as a potential qubit in the solid state is the negatively charged nitrogen-vacancy color center ($NV^-$) in diamond. It can be read-out optically ($\lambda \approx 637nm$), manipulated by microwave and initialized at room temperature [1-4]. In addition, quantum information encoded into these states presents extremely long coherence times ($T_1$) and ($T_2$) due to weak spin-orbit interaction and spin free host [2]. Indeed single and two-qubit operations with $NV^-$ centers in diamond were demonstrated by several groups [4]-[6]. The optimal configuration should minimize the dipole-dipole interaction between the optically active centers to avoid decoherence which would jeopardize the information storage i.e. the qubits should interact only via photons emitted by one and absorbed by another distant one. Hence the photons need to be transferred over macroscopic distances. This is best achieved by coupling the NV centers to a waveguide made of diamond. This approach has been adopted by several researchers who have designed and realized photonic crystals in diamond. The basic idea is that each qubit (the $NV^-$ site in diamond) is coupled to a high-Q cavity, and the cavities are optically interconnected via diamond waveguides [2],[7]-[9].

The first photonic crystal nanocavity was realized in nanodiamond [9], however, as nano diamond suffers from optical losses [9] and from broadening of $NV^-$ emission [10], single crystal diamond is a better material for the realization of quantum devices based on the $NV^-$ center. Major progress in diamond photonics and processing has been made in the last years, including the design of ultra-high-$Q$ cavity structures [11-14], demonstration single-photon sources [15] and the development of new patterning and material processing schemes [15,16]. Most of these are based on structures realized in thin diamond membranes. However, the performance of these is, as yet, unsatisfactory, and the realization of thin membranes with high optical quality, as required for the fabrication of photonic crystal structures is still a major challenge [16]. Therefore, a non-membrane based cavity design is highly desirable.

Here we describe a novel approach to the design of photonic crystals not based on the requirement of thin membrane. We propose nanobeam geometry capable to support a laterally confined 1D photonic crystal, and a high Q cavity in it. Nanobeams have drawn much attention in the last years and were implemented for various material systems, such as silicon [17,18], silicon nitride [18,20], silicon dioxide [21] and diamond [22]. However, most of these nanobeam designs required a pre-existing membrane [17-21] or relied on edge milling/relocation [22], producing a rectangular cross-section beam. An alternative fabrication technique on a bulk substrate is inclined etching (or milling) in such geometry that the material is cut in two opposite directions to form a beam with a triangular cross section (see

Fig.1-a). The etching (or milling) may result in a free standing structure. In the case of a diamond substrate, this process avoids the commonly used high energy ion implantation for the formation of a membrane as is the case in [16]. We have recently demonstrated the realization of a triangular cross section beam fabricated in a single crystal diamond with a rather low-$Q$ ($Q\approx300$) [22]. In the present work we show how to design an ultra-high-$Q$ nanobeam cavity in diamond applying the triangular beam approach. We explore the design parameter space and analyze the expected cavity mode characteristics.

In Section 2 we describe the triangular cross-section nanobeam geometry. We analyze the nanobeam waveguide dispersion curve and explain the mechanism of cavity mode confinement. In Section 3 we present the cavity mode profile and analyze the parameter space for implementation of a high $Q$ cavity. The highest quality factor and smallest mode volume obtained here are $Q\approx2.51\times10^6$ and $V_m=1.06\times(\lambda/n)^3$, respectively. The results and their implications for future applications are summarized in section 4.

## 2. The Triangular cross-section cavity geometry

We consider first a one-dimensional (1D) structure, in the form of a waveguide along the $z$-axis. The cross section of the waveguide is physically separated from the substrate (thus forming a "bridge" or a "beam") referred to below as a "nanobeam". The nano-beam profile in the $xy$ plane is triangular, having a height $H$ and a width $W$. Such a structure can be produced by inclined etching (or milling), as recently demonstrated by us in [22]. A grating along the $z$-direction consists of a periodic set of rectangular holes modulating (or perforating) the beam. The grating lattice constant is denoted as $a$, the hole widths in the $x$ and $z$ directions are $W_x$ and $W_z$, respectively (see Fig. 1a).

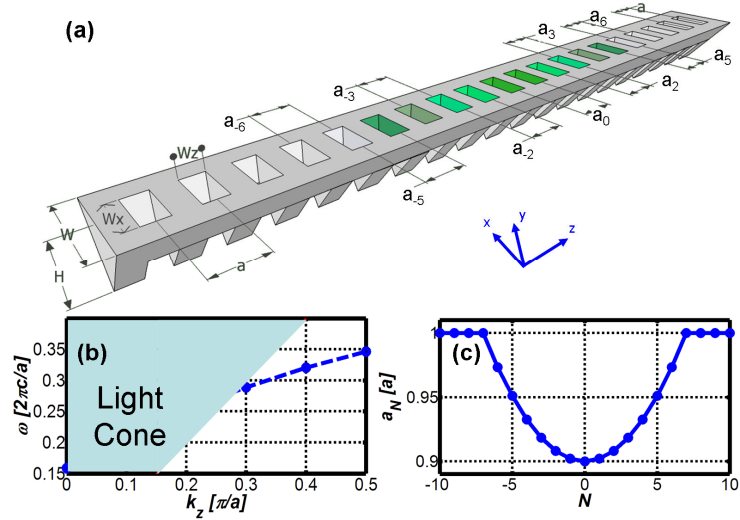

Fig. 1. The triangular cross section nanobeam. (a) The detailed geometry and notation of the cavity. $a_0$ denotes lattice constant in the defect center ($a_0=0.9a$), while $a_N$ denotes gradually increased lattice constant ($a_N\leq a$). (b) The dispersion curve $\omega$ vs. $k_z$ for a periodic waveguide structure in the 1st Irreducible Brillouin zone. The mode responsible for the cavity confinement is depicted. In blue the light cone area is shown. (c) The parabolic lattice constant variation responsible for the defect formation is presented ($a_N=0.9a+(N/7)^2\times0.1a$

This structure can conceptually be treated as *1D* photonic crystal (PC) waveguide. The electromagnetic mode profile is confined into this crystal by a total-internal-reflection (TIR) in the $xy$ plane, while its confinement/propagation in the $z$-direction is governed by the grating. Note that since the structure is not symmetric in the $y$-direction the modes can no longer be divided into *even/odd* groups along this axis, in contrast to PC in rectangular nanobeams where this is possible. The dispersion curve of the waveguide with a

*symmetric/anti-symmetric* boundary in the *x-z* directions is calculated by 3D Finite-Difference-Time-Domain (FDTD) with periodic conditions in the *z*-direction as shown in Fig. 1b. (Symmetric boundary conditions in the *x=0* direction implies that the magnetic field *y* component ($H_y$) satisfies $H_y(x,y,z)= H_y(-x,y,z)$, while the anti-symmetric condition satisfies $H_y(x,y,z)= -H_y(-x,y,z)$).

The cavity is formed by introducing a defect in the periodic structure (departure from periodicity along z) in a double hetero-structure approach, similar to that used for high *Q* cavities in planar photonic crystals [24]. The lattice constant is decreased in the defect region which is surrounded by the waveguides. By that, the mode eigen-frequency in the defect region is pushed upward with respect to that in the waveguide. This frequency separation provides mode confinement in the *z* direction, while the confinement in *xy* plane is governed by TIR. To delocalize the mode in *z*, thus increasing the localization in $k_z$, a gentle lattice variation is designed. The lattice in the defect region is increased parabolically from $a_0=0.9a$ at the defect center towards *a* in the waveguides regions, according to the formula: $a_N=0.9a+(N/7)^2 \times 0.1a$ *(for N≤7)* as suggested in [21].

The computations presented were performed by using a 3D-FDTD algorithm. The computational cell has dimensions: *[0,6a] ×[-2a,2a] ×[0,30a]* in the *x×y×z* directions. The discretization chosen for *x,y,z* directions was *25,25,50* points per *a*, respectively. This cell is surrounded by a perfectly matched layer (PML) and mirror (symmetric and anti-symmetric) boundaries (along *x=0* and *z=0*). The computation convergence for larger cell and better resolution has yielded similar *Q* values.

## 3 The cavity Parameter Space

We start with analyzing the cavity mode with the following geometry: *H=1.1a, W=3.82a, L=60a, $W_x$=0.5W, $W_z$=0.5a*. The cavity mode electric energy density profiles in various positions and planes are shown in Figs. 2(a-c) Since the field in the mid-height plane is no longer pure *TE* all *x,y,z* components of the electric and magnetic field appear together at mid beam height (*y=0*). As one can observe the energy is concentrated in the beam center slightly above the mid-height with a maximum at *(0,0.14a, 0)*. This is the location at which the atomic system (or NV⁻ center in diamond) must be positioned to achieve the maximal cavity coupling. Although all profiles in the figure are normalized the electric energy density ($U_E$) on the upper beam surface is by more that one order of magnitude larger than in the lower one, which suggests preferential upward radiation. As one can observe from Fig. 2b, the dominant side radiation in *x* direction is from the beam edge. (Note that $U_E$ is by one order of magnitude larger at *x=0*, than on the edge *x=1.91a*) These features are expected from the non symmetric geometry of the simulated triangular nano-beam. Generally, in the nanobeam geometry *Q* is determined by several dominant loss mechanisms as given by Eq. (1)

$$1/Q = 1/Q_x + 1/Q_{+y} + 1/Q_{-y} + 1/Q_z \qquad (1)$$

The losses in *x* and *z* directions are inversion symmetric. The $Q_x$ is defined by TIR, while the $Q_z$ is defined by Bragg reflection (PC). In contrast to [21], the losses in the vertical direction ($Q_{+y}$, $Q_{-y}$) are not equal and the ratio $Q_{+y}/Q_{-y}$ may vary from *1/5÷1/20*. In the present case of the triangular beam, $Q_{+y}$ is the dominant loss mechanism. From Fig. 2d it is evident that in the far-field the cavity mode presents a vertical lobe.

The calculations predict that a cavity with the parameters given in the table should have attractive properties, i.e. a quality factor of $Q=1.165 \times 10^6$, with the mode volume $V_m=1.26 \times (\lambda/n)^3$, the resonant wave length $a/\lambda \approx 0.3166$ and the corresponding Purcell coefficient $F_{cav}=(3/4\pi^2) \times (Q/V_m)=7.04 \times 10^4$.

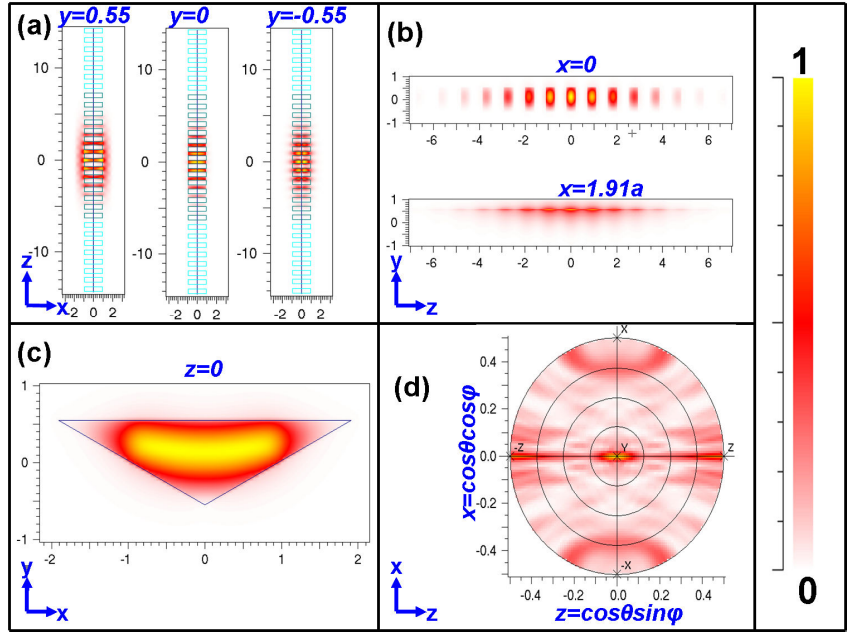

Fig. 2. Calculated cavity mode profiles of the normalized electric field energy density $U_E=\varepsilon/2|E|^2$. (a) $U_E$ in upper, mid and lower $y$ planes of the beam. (b) $U_E$ in the mid and edge $x$ planes of the beam. (c) $U_E$ in the $z=0$ plane. (d) Upper hemisphere far field distribution. The maximum appears at the vertical direction along the $y$ axis ($\theta=0$). All dimensions are in $a$ units.

In order to study how properties of the cavity $Q$, $V_m$ and $F_{cav}$ depend on the various geometries the dominant parameters $W_x$, $W$ and $H$ are changed, one at a time while keeping the others fixed. The results of these computations are shown in Fig. 3. Fig. 3a shows how $Q$, $V_m$ and $F_{cav}$ depend on $W_x$. As one can see the highest $Q$ ($Q=1.217\times10^6$) is obtained for $W_x=0.525W$ with $V_m=1.29\times(\lambda/n)^3$ and $F_{cav}=7.162\times10^4$.

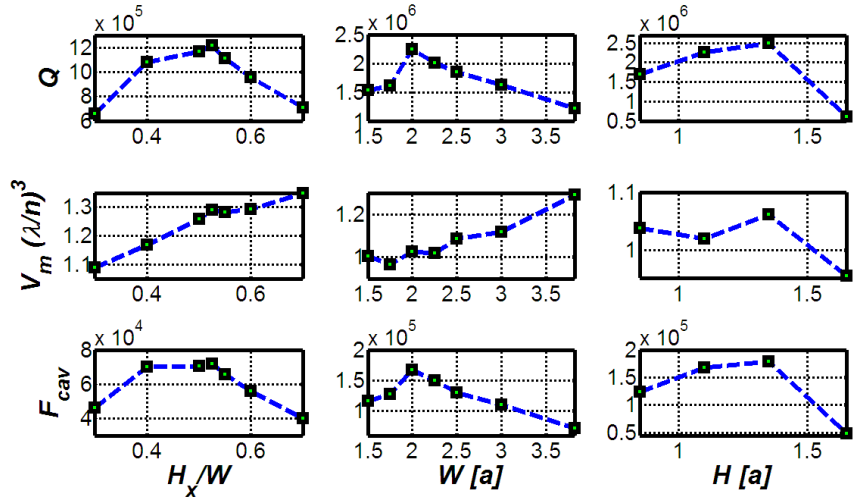

Fig. 3. Grating and Beam width variation: (a) $Q$, $V_m$, $F_{cav}$ dependence on $W_x/W$ when $W=3.82a$. (b) $Q$, $V_m$, $F_{cav}$ dependence on $W$ (in units of $a$) for $W_x=0.525W$. (c) $Q$, $V_m$, $F_{cav}$ dependence on $H$ (in units of $a$) for $H_x=0.525W$, $W=2a$. All other parameters are as described in Table 1.

The results obtained when the beam width $W$ is varied, while keeping the ratio $W_x/W=0.525$. are given in figure 3b showing that for $W=2a$ the maximal $Q$ is $Q=2.26\times10^6$, with $V_m=1.025\times(\lambda/n)^3$ and $F_{cav}\approx1.68\times10^5$. Note that $Q$ increase and $V_m$ decrease compared their values at $W=3.82a$. This leads to the improvement of $F_{cav}$ by a factor of ~2.35. Fig. 3c displays the dependence of the cavity properties on $H$ showing that for $H=1.35a$, the mode at the frequency $a/\lambda\approx0.339$ exhibits $Q=2.51\times10^6$, $V_m=1.062\times(\lambda/n)^3$ and $F_{cav}\approx1.79\times10^5$. The optimal cavity parameters are given in Table 1. This $Q$ value is ~2 times higher than that obtained for the PC slab analogue [14].

Table 1. The near optimal design parameters

| Beam Parameters | | | Waveguide Grating | |
|---|---|---|---|---|
| $H$ | $W$ | $L$ | $W_x$ | $W_z$ |
| 1.35a | 2a | 60a | 0.525W | 0.5a |

Finally, the dependence of Q on the separation between the substrate and the beam ($H_{sub}$) is explored. The results of these computations are given in Fig. 4. As one can observe, $H_{sub}=2.4a$ is sufficient to obtain a very high $Q$. In comparison a PC moderate-$Q$ design containing 3-missing holes with a similar separation yields a $Q$ value smaller by at least 2 orders of magnitude ( $Q\sim10^4$ ) [16]. The reason for the relatively small value required by the triangular beam geometry described here is the preferential upward radiation in this design, hence the need of a much lower device-substrate separation than for the nanobeam.

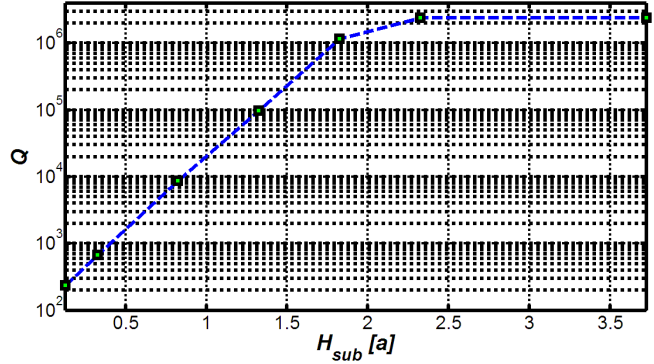

Fig. 4 Nanobeam $Q$ vs. distance to the substrate $H_{sub}$.

**4 Summary**

We have introduced a novel nanocavity design based on a triangular cross section nanobeam. This design is shown to have several advantages over the rectangular cross section based on a flat membrane geometry. In the case of diamond, the fabrication process of the triangular nanobeam avoids the detrimental effect of ion implantation damage to form a membrane as it relies on a tilted angle milling process. The proposed nanobeam cavity design is predicted to have an ultra high-$Q$ value with $Q=2.51\times10^6$, and a low mode volume of $V_m = 1.062\times(\lambda/n)^3$. The preferential upward radiation of the triangular nanobeam makes this geometry very attractive for the realization of vertical lasing lasers in III-V materials. The proposed design should be applicable not only to diamond but also to other materials, like sapphire, for which no readily available micro-machining technologies exist.

**Acknowledgement**
IB would like to thank Dr. Anne Weill-Zrahia, project manager of the Technion NANCO Computer Cluster for her devotion and indispensable help that made this work possible. Partial support of the Russell Berrie Nanotechnology Institute at the Technion and German Israeli Foundation (GIF) is acknowledged by all authors.